\documentclass[showpacs,amssymb,twocolumn,floatfix,aps,pre,notitlepage]{revtex4-2}

\usepackage{amssymb,amsfonts,amsmath,bm}
\usepackage{graphics,graphicx}
\usepackage{color}
 
\begin{document}

\title{Ising ferromagnets and antiferromagnets in an imaginary magnetic field}

\author{Roman Kr\v{c}m\'ar, Andrej Gendiar and Ladislav \v{S}amaj}
\affiliation{Institute of Physics, Slovak Academy of Sciences,
D\'ubravsk\'a cesta 9, 84511 Bratislava, Slovakia}

\begin{abstract}
We study classical Ising spin-$\frac{1}{2}$ models on the 2D square
lattice with ferromagnetic or antiferromagnetic nearest-neighbor interactions,
under the effect of a pure imaginary magnetic field.
The complex Boltzmann weights of spin configurations cannot be interpreted
as a probability distribution which prevents from application of standard
statistical algorithms.
In this work, the mapping of the Ising spin models under consideration onto
symmetric vertex models leads to real (positive or negative) Boltzmann weights.
This enables us to apply accurate numerical methods based on
the renormalization of the density matrix, namely the corner transfer matrix
renormalization group and the higher-order tensor renormalization group.
For the 2D antiferromagnet, varying the imaginary magnetic field
we calculate with a high accuracy the curve of critical points related to
the symmetry breaking of magnetizations on the interwoven sublattices.
The critical exponent $\beta$ and the anomaly number $c$ are shown to
be constant along the critical line, equal to their values $\beta=\frac{1}{8}$
and $c=\frac{1}{2}$ for the 2D Ising in a zero magnetic field. 
The 2D ferromagnets behave in analogy with their 1D counterparts defined on a
chain of sites, namely there exists a transient temperature which splits
the temperature range into its high-temperature and low-temperature parts.
The free energy and the magnetization are well defined in the high-temperature
region.  
In the low-temperature region, the free energy exhibits singularities at
the Yang-Lee zeros of the partition function and the magnetization is also
ill-defined: it varies chaotically with the size of the system.
The transient temperature is determined as a function of the imaginary
magnetic field by using the fact that from the high-temperature side
both the first derivative of the free energy with respect to the temperature
and the magnetization diverge at this temperature.
\end{abstract}


\date{\today} 

\maketitle

\section{Introduction}
Simulations of many systems in high-energy physics (QCD at finite
baryon density) and condensed matter (Hubbard model, antiferromagnetic
quantum spin chains, etc.) suffer from a severe sign problem.
For such systems, the complex Boltzmann weights of microscopic configurations
cannot be interpreted as a probability distribution which prevents from
application of standard statistical algorithms.

A prototype of systems with a severe sign problem in equilibrium statistical
mechanics is the Ising model of classical spins-$\frac{1}{2}$ on a lattice with
nearest-neighbor ferromagnetic or antiferromagnetic interactions,
under the effect of a pure imaginary magnetic field.
Similar generalizations in quantum mechanics are related to non-Hermitian
deformations of the transverse Ising quantum chains, e.g.,
via the inclusion of an imaginary longitudinal field
\cite{Uzelac80,Gehlen91,Deguchi09}.
The Ising model in an imaginary field was investigated mainly in its
antiferromagnetic version, on two-dimensional (2D) square or honeycomb
lattices.
The first studies were oriented to the location of Yang-Lee zeros of the
partition function in the complex magnetic field plane and Fisher zeros
in the complex temperature plane \cite{Matveev95,Matveev96,Matveev08,Kim04}.
Similarly as in the case of the real magnetic field
\cite{Muller77,Wu89,Kolesik93}, dividing the lattice into two interwoven
sublattices $A$ and $B$ the antiferromagnet exhibits two possible phases:
the disordered paramagnetic phase at high temperatures with equivalent
sublattice magnetizations $m_A=m_B$ and the symmetry-broken antiferromagnetic
phase at low temperatures with $m_A\ne m_B$.
As the imaginary magnetic field varies, the two phases are separated by a
curve of critical points.
This curve was specified with a good accuracy by computing the first
eight cumulants of the high-temperature expansion of the free energy
in Ref.~\cite{Azcoiti17}.
The mean-field analysis of the Ising antiferromagnet in an imaginary magnetic
field on a $D$-dimensional hypercubic lattice was carried out in
\cite{Azcoiti11}.

The Ising antiferromagnet in an imaginary magnetic field is exactly solvable
in one dimension (1D), see e.g. \cite{Azcoiti11,Azcoiti00}, where second-order
phase transitions are absent.
As concerns 2D, the exact solution of the antiferromagnetic Ising model
is known for zero field \cite{Onsager44} and for the (dimensionless) field
${\rm i}\pi/2$ \cite{Lee52}. 

The exact and numerical results obtained so far are restricted mainly to
the antiferromagnetic regime in 1D and 2D.
The ferromagnetic version of the model is not properly defined since it does
not correspond to a unitary theory for any value of the ferromagnetic
coupling \cite{Lee52,Azcoiti99}.
Consequently, the free energy and the magnetization per site of
the ferromagnet are not defined in the low-temperature region below
certain transient temperature \cite{Azcoiti99}.
Both in 1D and 2D, there exists a first-order phase transition from
the thermodynamically well-behaved high-temperature region to the
ill-defined low-temperature region at the transient temperature.
As is shown in this paper, the transient temperature can be detected from
the high-temperature side as either the divergence of the first derivative
of the free energy with respect to the coupling constant or the divergence
of the magnetization.

The aim of the present paper is twofold.
Firstly, for an arbitrary $D$-dimensional lattice, we construct a mapping of
the antiferromagnetic and ferromagnetic Ising models in an imaginary field
onto a symmetric vertex model whose local vertex (Boltzmann) weights are
real (positive or negative) numbers.
Secondly, the vertex representation of the original spin model permits us
to apply standard statistical methods, in particular accurate numerical
techniques based on the idea of renormalization group applied to the
density matrix.
For 2D antiferromagnets, varying the imaginary magnetic field
we calculate with a high accuracy the curve of critical points
and show the uniformity of the critical exponent $\beta$ and
the anomaly number $c$ along this curve.
As concerns 2D ferromagnets, the transient temperature below which
the free energy and the magnetization are not defined is determined as
a function of the imaginary magnetic field by using two different approaches.

The paper is organized as follows.
In Sec.~\ref{section:recapitulation}, we recapitulate briefly the 1D
exactly solvable case and discuss anomalies in the ferromagnetic version of
the Ising model in an imaginary field.
Sec.~\ref{section:mapping} deals with the mapping of the partition
function of the Ising model in an imaginary field onto the one of a symmetric
vertex model on the same lattice structure which exhibits real
(positive or negative) local Boltzmann vertex weights.
The mapping is constructed for both antiferromagnetic and ferromagnetic cases.
Applied numerical methods are described briefly in Sec.~\ref{section:methods};
some technicalities are moved to Appendix.
Numerical results for the critical properties of the 2D Ising
antiferromagnets are presented in Sec.~\ref{section:resultsantiferro}.
Numerical results for the 2D Ising ferromagnets are summarized in
Sec.~\ref{section:resultsferro}.
The emphasis is put on phenomena close to the first-order transition
temperature from the high-temperature to the low-temperature regimes. 
Sec.~\ref{section:conclusion} is a r\'esum\'e of the obtained results
with concluding remarks.

\section{Recapitulation of the 1D case} \label{section:recapitulation}
The 1D chain of $N$ Ising spins $\{ s_j=\pm 1\}_{j=1}^N$ with nearest-neighbor
couplings $J$ in a magnetic field $h$ is defined by the Hamiltonian
\begin{equation}
H = - J \sum_{j=1}^N s_j s_{j+1} - h \sum_{j=1}^N s_j ,
\end{equation}
with the cyclic boundary condition $s_{N+1}\equiv s_1$.
The partition function is given by
\begin{equation} \label{part}
Z_N = \sum_{\{ s\}} {\rm e}^{-\beta H} ,
\end{equation}
where $\beta=1/(k_{\rm B}T)$ is the inverse temperature and the summation
goes over all $2^N$ spin configurations.
Let us denote $\beta J\equiv F$ and consider the pure imaginary magnetic
field $\beta h\equiv {\rm i}\theta/2$.
Shifting $\theta$ by $2\pi$ induces for each vertex the same factor
${\rm e}^{\pm {\rm i}\pi} = -1$ which has no relevant effect on
the partition function (\ref{part}).
The partition function is also invariant with respect to the transformation
$\theta\to -\theta$ and therefore one can restrict oneself to
$\theta\in [0,\pi]$.
The $2\times 2$ transfer matrix
\begin{equation}
T = \begin{pmatrix}
{\rm e}^{F+{\rm i}\frac{\theta}{2}}&{\rm e}^{-F}\\
{\rm e}^{-F}&{\rm e}^{F-{\rm i}\frac{\theta}{2}}
\end{pmatrix}
\end{equation}  
has two eigenvalues of the form
\begin{equation} \label{eigenvalues}
\lambda_{\pm}(\theta) = {\rm e}^F \cos\left(\frac{\theta}{2}\right)
\pm \sqrt{{\rm e}^{-2F}-{\rm e}^{2F}\sin^2\left(\frac{\theta}{2}\right)} .  
\end{equation}
The partition function (\ref{part}) is determined by the eigenvalues of
the transfer matrix as follows
\begin{equation} \label{partition}
Z_N = \lambda_+^N + \lambda_-^N .
\end{equation}
The free energy per spin $f$ is defined by
\begin{equation} \label{freeenergy}
- \beta f_N = \frac{1}{N} \ln Z_N
\end{equation}
and the magnetization per spin $m_N = \langle s_j \rangle$,
which in 1D does not depend on the site index $j=1,2,\ldots$, by
\begin{equation} \label{magnetization}
m_N = - \frac{\partial}{\partial \beta h} \beta f_N
= 2 {\rm i} \frac{\partial}{\partial \theta} \beta f_N . 
\end{equation}
Note that the magnetization $m_N$, which is bounded by $0< \vert m \vert \le 1$
for real magnetic fields, can have magnitude larger than $1$ for
imaginary magnetic fields.

The spin system is usually studied in the thermodynamic limit $N\to \infty$.
The analysis of the above equations depends on whether the (dimensionless)
coupling constant $F$ is positive (ferromagnet) or negative (antiferromagnet).

\subsection{1D antiferromagnet} \label{1Dantiferro}
If $F<0$, the argument of the square root in (\ref{eigenvalues}) is always
positive which implies real eigenvalues $\lambda_{\pm}$, $\lambda_+>0$ and
$\lambda_-<0$; since $\lambda_+>\vert \lambda_-\vert$ the partition function
(\ref{partition}) is real and positive.
In the limit $N\to\infty$, from the two summands in (\ref{partition})
$\lambda_+^N$ dominates, so that the free energy per site
$f=\lim_{N\to\infty} f_N$ is given by
\begin{equation} \label{betaf}
- \beta f = F + \ln\left[ \cos\left(\frac{\theta}{2}\right) +
\sqrt{{\rm e}^{-4F}-\sin^2\left(\frac{\theta}{2}\right)} \right] .   
\end{equation}
The magnetization per site $m=\lim_{N\to\infty} m_N$ is given by  
\begin{equation} \label{m}
- {\rm i} m = \frac{\sin\left(\frac{\theta}{2}\right)}{
\sqrt{{\rm e}^{-4F}-\sin^2\left(\frac{\theta}{2}\right)}} .
\end{equation}
There is no second-order phase transition in 1D.

\subsection{1D ferromagnet} \label{1Dferro}
If $F>0$, the argument of the square root in (\ref{eigenvalues})
can have both positive and negative signs.
For a fixed value of $\theta\in [0,\pi]$, let us introduce a ``transition''
coupling $F^*(\theta)$, 
\begin{equation} \label{defFstar}
{\rm e}^{-2 F^*} = \sin\left(\frac{\theta}{2}\right) , 
\end{equation}
at which the argument of the square root in (\ref{eigenvalues}) vanishes.

In the high-temperature region $0<F<F^*$, the argument of the square root
is positive which implies that $\lambda_+>\vert \lambda_-\vert$ and
one can use the previous formulas (\ref{betaf}) and (\ref{m}).
Note that as $F$ approaches $F^*$ the magnetization (\ref{m}) diverges. 

In the low-temperature region $F>F^*$, the argument of the square root
is negative and, consequently, the complex conjugate eigenvalues
\begin{equation} \label{eigenvaluesferro}
\lambda_{\pm}(\theta) = {\rm e}^F \cos\left(\frac{\theta}{2}\right) \pm {\rm i}
\sqrt{{\rm e}^{2F}\sin^2\left(\frac{\theta}{2}\right)-{\rm e}^{-2F}}  
\end{equation}
have in polar coordinates the same modulus and the opposite phases:
\begin{equation}
\lambda_{\pm}(\theta) = \sqrt{{\rm e}^{2F}-{\rm e}^{-2F}}
\exp\left( \pm {\rm i}\varphi \right) ,  
\end{equation}
where
\begin{equation}
\varphi(F,\theta) = \arccos  \left[
\frac{\cos\left(\frac{\theta}{2}\right)}{\sqrt{1-{\rm e}^{-4 F}}} \right] .    
\end{equation}

The partition function (\ref{partition}) reads as
\begin{equation}
Z_N = 2 \left( {\rm e}^{2F} - {\rm e}^{-2F} \right)^{N/2} \cos(N \varphi) ,
\end{equation}
The Yang-Lee zeros of the partition function thus exist exclusively 
in the low-temperature region $F>F^*$ and correspond to
the following irrational values of $\varphi$:
\begin{equation} \label{YL}
\varphi(F,\theta) = \frac{2j-1}{2 N} \pi , \qquad
j = 1, 2,\ldots, N . 
\end{equation}  
Since $Z_N=0$, the free energy goes to $-\infty$ at these points
which become dense in the limit $N\to\infty$.
This means that the free energy is not defined for $F>F^*$.

As is shown in this paragraph, it is a mathematical curiosity that if one
fixes the value of $\varphi$ outside of the Yang-Lee set (\ref{YL}), say
$\varphi$ is a rational number, the expression for the free energy converges
when increasing the number of sites $N\to\infty$.
The term $\cos(\varphi N)$ will change its sign with increasing $N$. 
The oscillating sign of the partition function does not represent any
problem in the definition of the free energy per spin (\ref{freeenergy})
since the principal value of the complex logarithm $\ln (-1) = {\rm i}\pi$,
when divided by $N$, goes to 0 in the limit $N\to\infty$.
Taking the absolute value of the partition function in the definition of
the free energy per spin (\ref{freeenergy}), one gets
\begin{equation}
-\beta f_N = \frac{1}{2} \ln \left( {\rm e}^{2F}-{\rm e}^{-2F} \right)
+ \frac{1}{N} \ln \vert 2 \cos(N \varphi) \vert .  
\end{equation}
Using the formula \cite{Gradshteyn}
\begin{equation}
2 \cos ( N \varphi) = 2^N \prod_{k=1}^N
\sin \left( \varphi + \frac{2k-1}{2 N} \pi \right) ,
\end{equation}  
one obtains that
\begin{equation} \label{discrete}
\frac{1}{N} \ln \vert 2 \cos ( N \varphi) \vert = \ln 2
+ \frac{1}{N} \sum_{k=1}^N \ln \left\vert
\sin \left( \varphi + \frac{2k-1}{2N} \pi \right) \right\vert .
\end{equation}  
According to the Euler-Maclaurin formula \cite{EM}
\begin{eqnarray}
\sum_{n=a}^b f(n) & \sim & \int_a^b {\rm d}x\, f(x)
+ \frac{1}{2} \left[ f(a) + f(b) \right] \nonumber \\
& & + \sum_{k=1}^{\infty} \frac{B_{2k}}{(2 k)!} \left[
f^{(2k-1)}(b) - f^{(2k-1)}(a) \right]
\end{eqnarray}
with $a, b$ being integers and $\{ B_{2k} \}$ the Bernoulli numbers,
the discrete sum on the rhs of (\ref{discrete}) is nothing but
a Riemann integral plus large-$N$ corrections.
The set of Yang-Lee zeros (\ref{YL}) becomes dense in the thermodynamic
limit $N\to\infty$.
If $\varphi$ belongs to the set of Yang-Lee zeros (\ref{YL}),
the continualization of (\ref{discrete}) is not possible as one of
the summands, namely the one with $j+k-1=N$, diverges.
When $\varphi$ does not belong to the set of Yang-Lee zeros (\ref{YL}),
say it is a rational number at an infinitesimal distance $1/N$ from
Yang-Lee zeros in its neighborhood, the problematic summand in (\ref{discrete})
is of order $\ln[\sin(1/N)]/N \sim - (\ln N)/N$ and vanishes in the limit
$N\to\infty$, so that
\begin{equation} \label{limit}
\frac{1}{N} \ln \vert 2 \cos ( N \varphi) \vert = \ln 2 + \int_0^1 {\rm d}t\,
\ln \left\vert \sin(\varphi + \pi t) \right\vert + o(1) . 
\end{equation}
The integral over $t$ exactly cancels the term $\ln 2$ for any
value of $\varphi$.
We then suggest that as soon as $\varphi$ is a rational number, the
thermodynamic $N\to\infty$ limit of the lhs of Eq. (\ref{limit}) exists
and equals to $0$; one can check this suggestion numerically by fixing
$\varphi$ (say to an integer) and going with $N$ to extremely large values.
The thermodynamic limit of the free energy then reads as
\begin{equation} \label{freee}
-\beta f = \frac{1}{2} \ln \left( {\rm e}^{2F}-{\rm e}^{-2F} \right) .
\end{equation}
Thus, excluding from the consideration the set of Yang-Lee zeros (\ref{YL}),
the free energy is a continuous function of $F$ when passing through
the point $F=F^*$ as it should be; this can be seen by inserting
$\cos(\theta/2) = \sqrt{1-{\rm e}^{-4F^*}}$ into (\ref{betaf}) taking
in the limit $F\to {F^*}^-$ and comparing to (\ref{freee})
taking in the limit $F\to {F^*}^+$.
As concerns the derivative of the free energy with respect to the coupling $F$,
it diverges for $F\to {F^*}^-$ and converges to a finite number when
$F\to {F^*}^+$ which signalizes a first-order phase transition
at the transient point $F^*$.
It should be emphasized that as the set of Yang-Lee zeros (\ref{YL})
is dense in the limit $N\to\infty$, the above mathematical analysis is of
limited physical interest.

The magnetization per site (\ref{magnetization})
\begin{equation}
m_N = 2 {\rm i} \frac{\partial\varphi}{\partial\theta} \tan(\varphi N) 
\end{equation}
oscillates with increasing $N$ and so it does not exhibit a well defined
thermodynamic limit in the low-temperature region $F>F^*$.

\section{Mapping onto a symmetric vertex model} \label{section:mapping}
In this section, we consider a spin-$\frac{1}{2}$ Ising model on
a general $D$-dimensional lattice structure with coordination number
$q=2,3,\ldots$.
The spin Hamiltonian $H$ is given by  
\begin{equation} \label{Hamiltonian}
-\beta H = F \sum_{\langle j,k\rangle} s_j s_k + {\rm i} \frac{\theta}{2}
\sum_j s_j ,  
\end{equation}  
where the first sum goes over all nearest-neighbor pairs of lattice sites
and the second sum over all lattice sites.
The partition function is defined by (\ref{part}).

In Ising systems, microscopic spins $s=\pm 1$ are associated with
lattice sites and the nearest-neighbor spins interact along
edges connecting the nearest-neighbor vertices.
In two-state vertex models, microscopic states $\sigma = \pm 1$
are attached to the edges of the lattice.
For a given ``global'' configuration of edge states, every vertex sees
a ``local'' configuration of edge states with the corresponding Boltzmann
vertex weight.
The partition function of the vertex system is defined by
\begin{equation}
Z = \sum_{\{\sigma\}} \prod ({\rm weights}) ,
\end{equation}
where the sum goes over all configurations of edge states and the product
is over all vertex weights in the lattice.
A special case of two-state vertex systems is a \emph{symmetric} vertex model
whose local vertex weights depend only on the number of incident edges
in, say, $(-)$ state; in other words, for a vertex, any permutation of edge
states in space leaves the local vertex weight invariant.

Every system of Ising spins on a lattice can be mapped onto a symmetric
two-state vertex model formulated on the same lattice structure by using
mapping methods \cite{Kolesik93,Samaj92} based on a gauge transformation
\cite{Wegner73} which represents a generalization of the duality
transformation and the weak-graph expansion \cite{Nagle68}.
The mapping depends on whether the spin coupling $F$ is
ferromagnetic or antiferromagnetic. 

\subsection{Ising antiferromagnet} \label{section:mappinga}
In the case of an antiferromagnetic coupling $F<0$ it holds that
$F=-\vert F\vert$.
The Ising model on a lattice with the coordination number $q$ can be
represented as a vertex system when one decorates each edge by
a new two-coordinated vertex and attach to line fragments
two-state variables $\sigma=\pm 1$.
To reproduce the partition function of the Ising model, one attaches to
the new decoration vertices the $2\times 2$ interaction matrix
\begin{equation} \label{interaction}
{\bf V} \equiv \begin{pmatrix}
V_{+,+} & V_{+,-} \\
V_{-,+} & V_{-,-}
\end{pmatrix}
= \begin{pmatrix}
{\rm e}^F&{\rm e}^{-F}\\
{\rm e}^{-F}&{\rm e}^F
\end{pmatrix}
\end{equation}  
and to the vertices of the original lattice with a local configuration
of adjacent edges $\{ \sigma_1,\sigma_2,\ldots,\sigma_q \}$
the vertex weights
\begin{eqnarray}
v(\sigma_1,\sigma_2,\ldots,\sigma_q) & = & {\rm e}^{{\rm i}\frac{\theta}{2}}
\delta(\sigma_1,+) \delta(\sigma_2,+) \cdots \delta(\sigma_q,+) \nonumber \\
& & + {\rm e}^{-{\rm i}\frac{\theta}{2}}
\delta(\sigma_1,-) \delta(\sigma_2,-) \cdots \delta(\sigma_q,-) . \nonumber \\
& & 
\end{eqnarray}
In this way, two admissible configurations around a vertex on
the original lattice, all adjacent edges in the same either $(+)$ or
$(-)$ state, are identified with the $(+)$ or $(-)$ state of the spin on
that vertex.
For a given edge composed of two line fragments in states $\sigma'$ and
$\sigma''$, the contribution to the partition function can be schematically
expressed as
\begin{equation} \label{schema}
\sum_{\sigma',\sigma''} v(\ldots,\sigma',\ldots) V_{\sigma',\sigma''}
v(\ldots,\sigma'',\ldots) .
\end{equation}
The interaction matrix (\ref{interaction}) can be written as product
of a matrix ${\bf W}$ and its transpose ${\bf W}^{\rm T}$ in many ways;
let us apply the following factorization
\begin{equation} \label{Wmatrix}
{\bf V} = {\bf W} {\bf W}^{\rm T} , \qquad
{\bf W} = \begin{pmatrix}
\sqrt{\cosh F} & {\rm i} \sqrt{\sinh \vert F\vert} \\
\sqrt{\cosh F} & - {\rm i} \sqrt{\sinh \vert F\vert}
\end{pmatrix} . 
\end{equation}
The next step is to use the relation
$V_{\sigma',\sigma''} = \sum_{\sigma} W_{\sigma',\sigma} W_{\sigma'',\sigma}$
in (\ref{schema}) to eliminate the decoration vertices by attaching ${\bf W}$
to the left endpoint and ${\bf W}^{\rm T}$ to the right endpoint of each edge.
In this way one obtains the pure two-state vertex model on the original
lattice structure defined by the vertex weights
\begin{eqnarray}
w(\sigma_1,\sigma_2,\ldots,\sigma_q)
& = & \sum_{\sigma'_1,\sigma'_2,\ldots,\sigma'_q=\pm} 
v(\sigma'_1,\sigma'_2,\ldots,\sigma'_q) \nonumber \\ 
& & \times W_{\sigma'_1,\sigma_1} W_{\sigma'_2,\sigma_2} \cdots W_{\sigma'_q,\sigma_q} .
\end{eqnarray}  
Explicitly,
\begin{eqnarray}
w(\sigma_1,\sigma_2,\ldots,\sigma_q) & = &
{\rm e}^{{\rm i}\frac{\theta}{2}} W_{+,\sigma_1} W_{+,\sigma_2} \cdots W_{+,\sigma_q} 
\nonumber \\ & & +
{\rm e}^{-{\rm i}\frac{\theta}{2}} W_{-,\sigma_1} W_{-,\sigma_2} \cdots W_{-,\sigma_q} .
\phantom{aaa} \label{permutation}
\end{eqnarray}  
These vertex weights are invariant with respect to any permutation of edge
states and therefore they correspond to a symmetric vertex model.
It stands to reason that the partition function of the original Ising model
is identical by construction to the one of the symmetric vertex model on
the same lattice structure.

Let for the resulting symmetric vertex model $w_n$ $(n=0,1,\ldots,q)$ be
the vertex weight of edge configurations with $n$ adjacent edges in state $(-)$
and the remaining $q-n$ adjacent edges in state $(+)$.
Then, according to (\ref{permutation}), one has
\begin{equation}
w_n = {\rm e}^{{\rm i}\frac{\theta}{2}} W^n_{+,-} W^{q-n}_{+,+} +
{\rm e}^{-{\rm i}\frac{\theta}{2}} W^n_{-,-} W^{q-n}_{-,+} .
\end{equation}
According to the form of the ${\bf W}$-matrix (\ref{Wmatrix}),
the elements $W_{++}=W_{-+}=\sqrt{\cosh F}$ and
$W_{+-} = - W_{--} = {\rm i} \sqrt{\sinh F}$, so that
\begin{equation}
w_n = \left( \cosh F \right)^{\frac{q-n}{2}}
\left( \sinh \vert F\vert \right)^{\frac{n}{2}} {\rm i}^n 
\left[ {\rm e}^{{\rm i}\frac{\theta}{2}} + (-1)^n {\rm e}^{-{\rm i}\frac{\theta}{2}} 
\right] . 
\end{equation}
For even number $n$ of adjacent edges in state $(-)$, it holds that
${\rm i}^n=(-1)^{\frac{n}{2}}$, $(-1)^n=1$ and the consequent sum of
the exponentials in the square brackets results in
$2 \cos\left(\frac{\theta}{2} \right)$, i.e.,
\begin{equation}
w_n= 2 (-1)^{\frac{n}{2}} \left( \cosh F \right)^{\frac{q-n}{2}} \left(
\sinh \vert F\vert \right)^{\frac{n}{2}} \cos\left(\frac{\theta}{2} \right) .
\end{equation}
For odd $n$, it holds that ${\rm i}^n=(-1)^{\frac{n+1}{2}}/{\rm i}$, $(-1)^n=-1$
and the consequent difference of the exponentials in the square brackets,
divided by ${\rm i}$, results in $2 \sin\left(\frac{\theta}{2} \right)$, i.e.,
\begin{equation}
w_n= 2 (-1)^{\frac{n+1}{2}} \left( \cosh F \right)^{\frac{q-n}{2}} \left(
\sinh \vert F\vert \right)^{\frac{n}{2}} \sin\left(\frac{\theta}{2} \right) .
\end{equation}
We conclude that in the vertex picture all local Boltzmann weights are real,
positive or negative, as was needed.

\subsection{Ising ferromagnet} \label{section:mappingb}
To construct the mapping for the ferromagnetic Ising model, one
has to divide the lattice into two interwoven sublattices $A$ and $B$
and to change signs of spin variables $s_j\to -s_j$ at vertices of
say the $B$-sublattice.
This transformation has no effect on the partition function which is
the sum over all spin configurations.
On the other hand, the spin Hamiltonian (\ref{Hamiltonian}) is changed to
\begin{equation} \label{Hamiltoniannew}
-\beta H = - F \sum_{\langle j,k\rangle} s_j s_k + {\rm i} \frac{\theta}{2}
\sum_{j\in A} s_j - {\rm i} \frac{\theta}{2} \sum_{j\in B} s_j ,    
\end{equation}  
i.e., the ferromagnetic coupling $F>0$ is changed to the antiferromagnetic
one $-F<0$ and the sign of the imaginary magnetic field alternates
with the $A$ and $B$ sublattices. 
Having the antiferromagnetic coupling one can proceed as in the previous
subsection.
After the mapping, the vertex weights of the symmetric vertex model
depend on whether the vertex is on the sublattice $A$ or $B$.
If the vertex lies on the $A$-sublattice, the vertex weights are given by
\begin{equation}
w^{(A)}_n= 2 (-1)^{\frac{n}{2}} \left( \cosh F \right)^{\frac{q-n}{2}} \left(
\sinh F \right)^{\frac{n}{2}} \cos\left(\frac{\theta}{2} \right)
\end{equation}
for even $n$ and
\begin{equation}
w_n^{(A)}= 2 (-1)^{\frac{n+1}{2}} \left( \cosh F \right)^{\frac{q-n}{2}} \left(
\sinh F \right)^{\frac{n}{2}} \sin\left(\frac{\theta}{2} \right)
\end{equation}  
for odd $n$.
If the vertex lies on the $B$-sublattice, the vertex weights are given by
\begin{equation}
w^{(B)}_n= 2 (-1)^{\frac{n}{2}} \left( \cosh F \right)^{\frac{q-n}{2}} \left(
\sinh F \right)^{\frac{n}{2}} \cos\left(\frac{\theta}{2} \right)
\end{equation}
for even $n$ and
\begin{equation}
w_n^{(B)}= 2 (-1)^{\frac{n-1}{2}} \left( \cosh F \right)^{\frac{q-n}{2}} \left(
\sinh F \right)^{\frac{n}{2}} \sin\left(\frac{\theta}{2} \right)
\end{equation}  
for odd $n$.

\section{Numerical methods} \label{section:methods}
We apply two distinct numerical methods to the Ising models, which originate
in the density-matrix renormalization \cite{White92,White93,Schollwock05}.
Namely, we use the corner transfer matrix renormalization group
(CTMRG) \cite{Nishino96,Nishino97,Ueda05} and the higher-order tensor
renormalization group (HOTRG) \cite{Xie12} methods for that purpose.

1) The CTMRG method comes from the Baxter's corner-transfer-matrix approach,
originally proposed for the square-lattice Ising spins \cite{Baxterbook}.
CTMRG is used in this work to evaluate the von Neumann entropy
\begin{equation}
S = - {\rm Tr}\left(\rho \ln \rho\right) .
\end{equation}  
Here, $\rho$ represents a reduced density matrix, which is used for
the construction of the renormalization transformations.
At $\theta=0$, the Ising critical point is
$F=F_c=\frac{1}{2}\ln\left(1+\sqrt{2}\right)$ \cite{Onsager44}, in which
the von Neumann entropy $S$ logarithmically diverges with respect to
number of the spins $N$ \cite{Calabrese04,Ercolessi10}
\begin{equation} \label{entropyscale}
S \sim \frac{c}{12} \ln N, \qquad F=F_c.
\end{equation}
The parameter $c$ is an anomaly number (also known as the central charge)
determining the universality class of the statistical system.
For the 2D Ising model at zero magnetic field, the anomaly number
$c=\frac{1}{2}$.
We first evaluate the $N$-dependence of an effective anomaly number
at criticality $F_c$,
\begin{equation} \label{ceff}
c_{\rm eff}(N) = 12 \frac{\partial S}{\partial \ln N}.
\end{equation}
Finally, the asymptotic value of $c_{\rm eff}(N)$ yields
\begin{equation}
c = \lim_{N\to\infty} c_{\rm eff}(N) .
\end{equation}  

2) The HOTRG method is applied to the calculation of the free energy
(for both the antiferromagnet and the ferromagnet) in standard way.
In the symmetry broken phase ($F>F_c$) of the Ising antiferromagnet,
the magnetizations per spin $m_A$ and $m_B$, associated with the
two sublattices $A$ and $B$, respectively, differ.
Hence, nonzero magnetization difference results in
\begin{equation} \label{mminus}
m_{AB} = m_A - m_B \ne 0 .
\end{equation}  
Since it is not straightforward how to evaluate the imaginary magnetization
$m_{AB}$ by HOTRG, we proposed an {\it extended} impurity tensor $T_{AB}$
in order to distinguish the symmetry broken phase from the disordered one.
It is so because there is no concise way of how to observe the $Z_2$ broken
symmetry, provided that the real (non-imaginary) character of HOTRG tensors
has to be preserved. Further details of constructing the impurity-tensor
are briefly described in Appendix. 

\section{Numerical results for the 2D antiferromagnet}
\label{section:resultsantiferro}

\begin{figure}[t]
\begin{center}
\includegraphics[width=0.48\textwidth,clip]{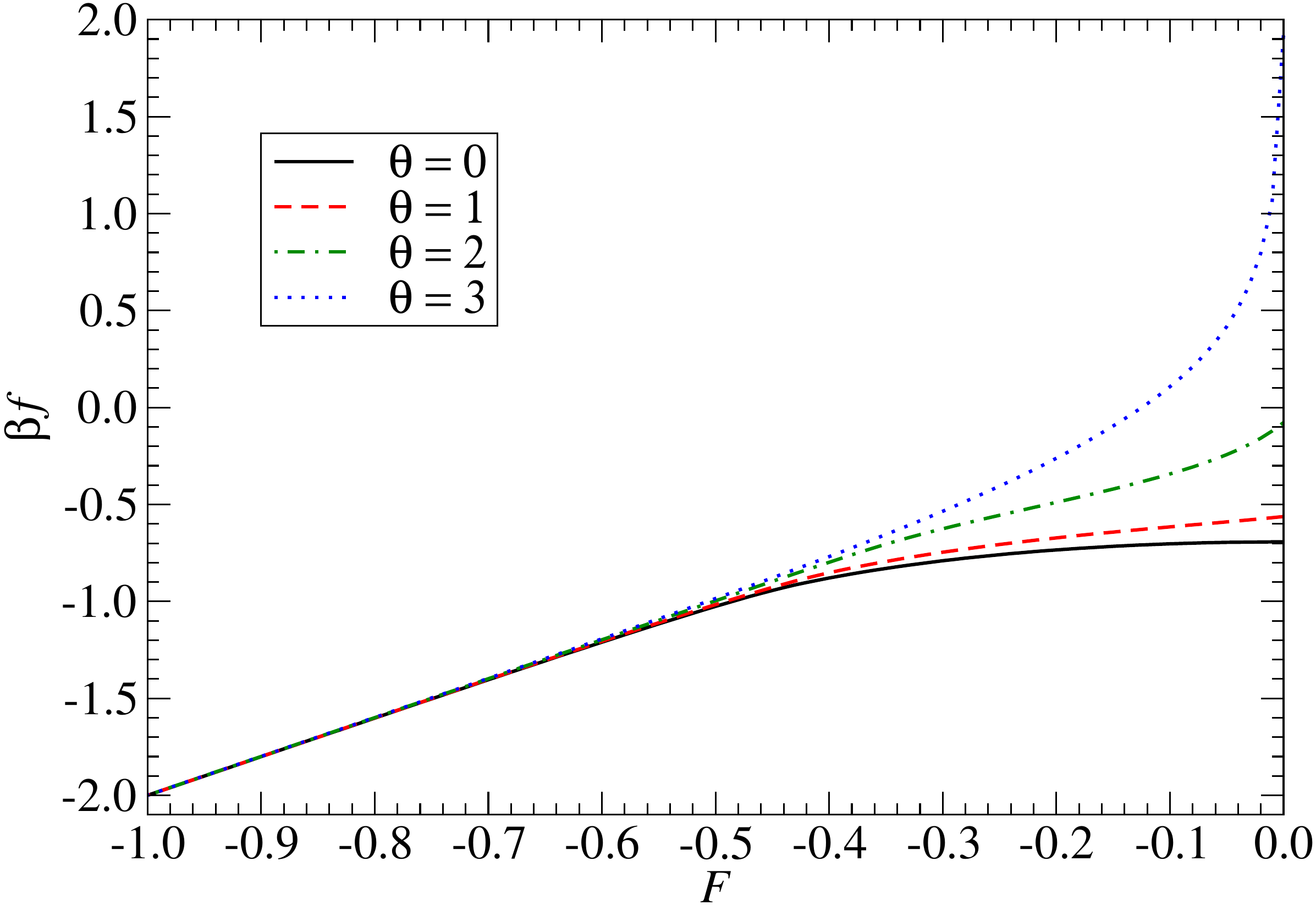}
\caption{The (dimensionless) free energy per spin $\beta f$ of
the 2D antiferromagnetic Ising model versus the coupling $F\le 0$ for
four values of the (dimensionless) imaginary magnetic field
$\beta H = {\rm i}\theta/2$: zero field $\theta=0$ (solid curve), $\theta=1$
(dashed curve), $\theta=2$ (dash-dotted curve) and $\theta=3$ (dotted curve).}
\label{fig:feantiferro} 
\end{center}
\end{figure}

Using the vertex representation of the 2D Ising antiferromagnet on the
square lattice $(q=4)$ derived in Sec.~\ref{section:mappinga},
the dependence of the (dimensionless) free energy per spin $\beta f$
on the coupling $F\le 0$ is pictured in Fig.~\ref{fig:feantiferro} for
the zero magnetic field $\theta=0$ (solid curve) and three values of
the imaginary magnetic field $\theta=1$ (dashed curve),
$\theta=2$ (dash-dotted curve) and $\theta=3$ (dotted curve);
the same notation will be used in what follows.
The spins become uncoupled in the limit $F\to 0$, so the curves end up at
the points $\beta f = - \ln \left[ 2\cos(\theta/2)\right]$.
The free energy is always an increasing function of the coupling $F$.
In the low-temperature region, for small enough antiferromagnetic
coupling $F\lessapprox -0.5$, the curves approach close to each other.
An analogous behavior is observed also in 1D.

\begin{figure}[t]
\begin{center}
\includegraphics[width=0.48\textwidth,clip]{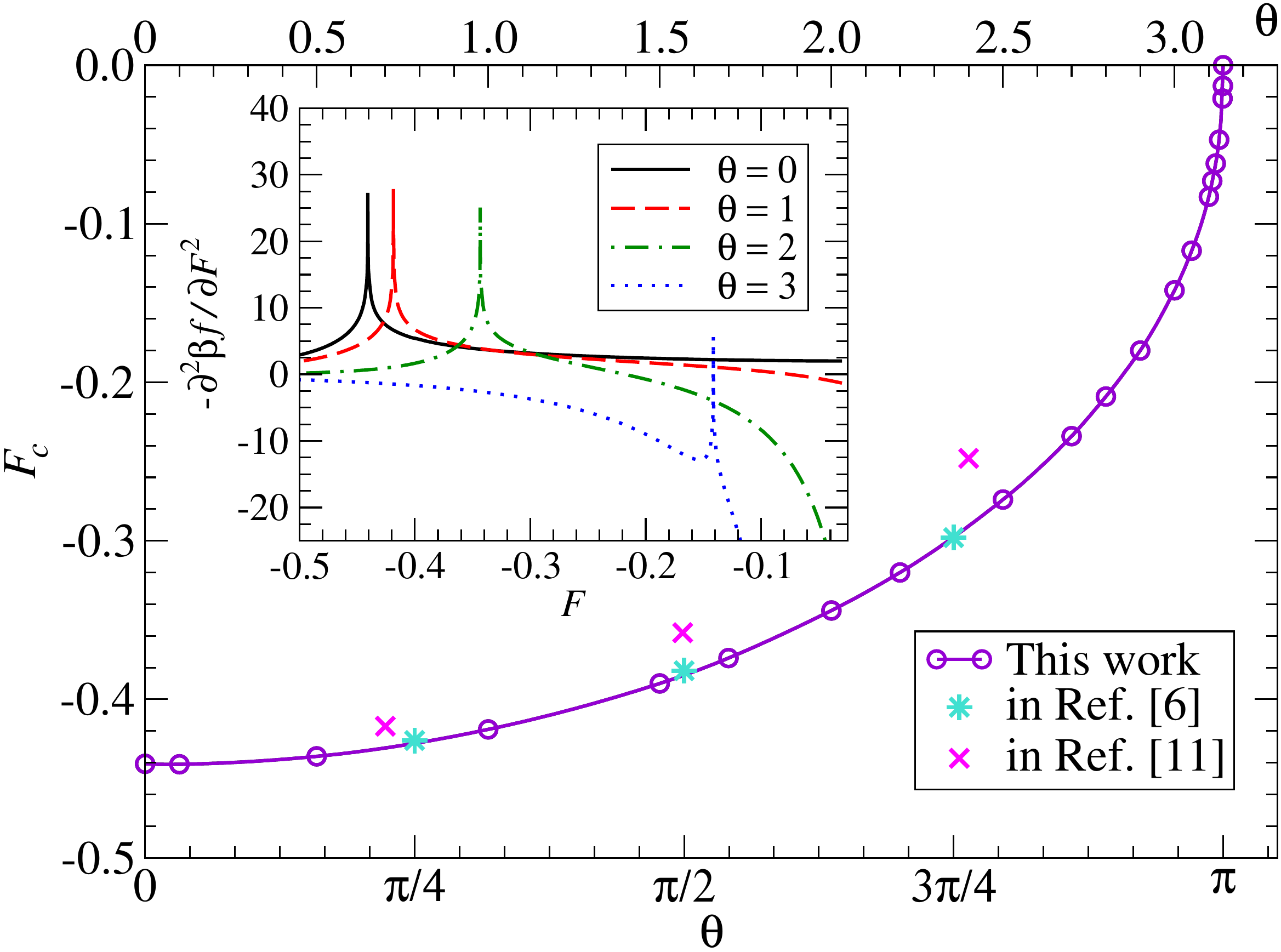}
\caption{Antiferromagnet:
The value of the critical coupling $F_c$ as the function of
the imaginary magnetic field $\theta\in [0,\pi]$.
The present data (open circles) are compared with the ones of
Ref.~\cite{Matveev08} (stars) and Ref.~\cite{Azcoiti17} (crosses).
The $F$-dependence of the second derivative of the free energy $\beta f$
with respect to $F$ is pictured in the inset for four
values $0,1,2,3$ of the imaginary magnetic field $\theta$;
the cusp divergence of the second derivative determines
the critical point $F_c(\theta)$.}
\label{fig:critF} 
\end{center}
\end{figure}

The phenomenon which does not occur in 1D is seen in the inset of
Fig.~\ref{fig:critF} where the $F$-dependence of the second derivative of
the free energy $\beta f$ with respect to $F$ is represented
for four values $0,1,2,3$ of the imaginary magnetic field $\theta$.
For each $\theta$, there is a critical point $F_c$ at which
the second derivative goes to $-\infty$.
The dependence of critical points $F_c$ on the imaginary magnetic field
$\theta\in [0,\pi]$ is pictured in the main body of Figure \ref{fig:critF}.
The present data (open circles) are compared with numerical data
from other works.  
The data of Ref.~\cite{Matveev08} (stars), obtained by calculating
complex-temperature zeros of the partition function for finite lattices
of relatively small sizes, are in a good agreement with our data.
The data of Ref.~\cite{Azcoiti17} (crosses), obtained by extrapolation
of the high-temperature cumulant expansion of the free energy into the
critical region, deviate much more from our data.
This is caused by the fact that only the first eight cumulants were taken
into account.
The numerical estimate of $F_c\approx -0.4410$ for
the zero magnetic field $\theta=0$ is in good agreement with the exact value
$F_c = -0.44068679\ldots$ obtained by Onsager \cite{Onsager44}.
The exact value $F_c=0$ for $\theta=\pi$ \cite{Lee52} is also reproduced
by our numerical calculations.
Note that a similar curve of critical points occurs for the 2D Ising
antiferromagnet in {\em real} non-zero magnetic fields \cite{Wu89,Kolesik93}.

\begin{figure}[t]
\begin{center}
\includegraphics[width=0.48\textwidth,clip]{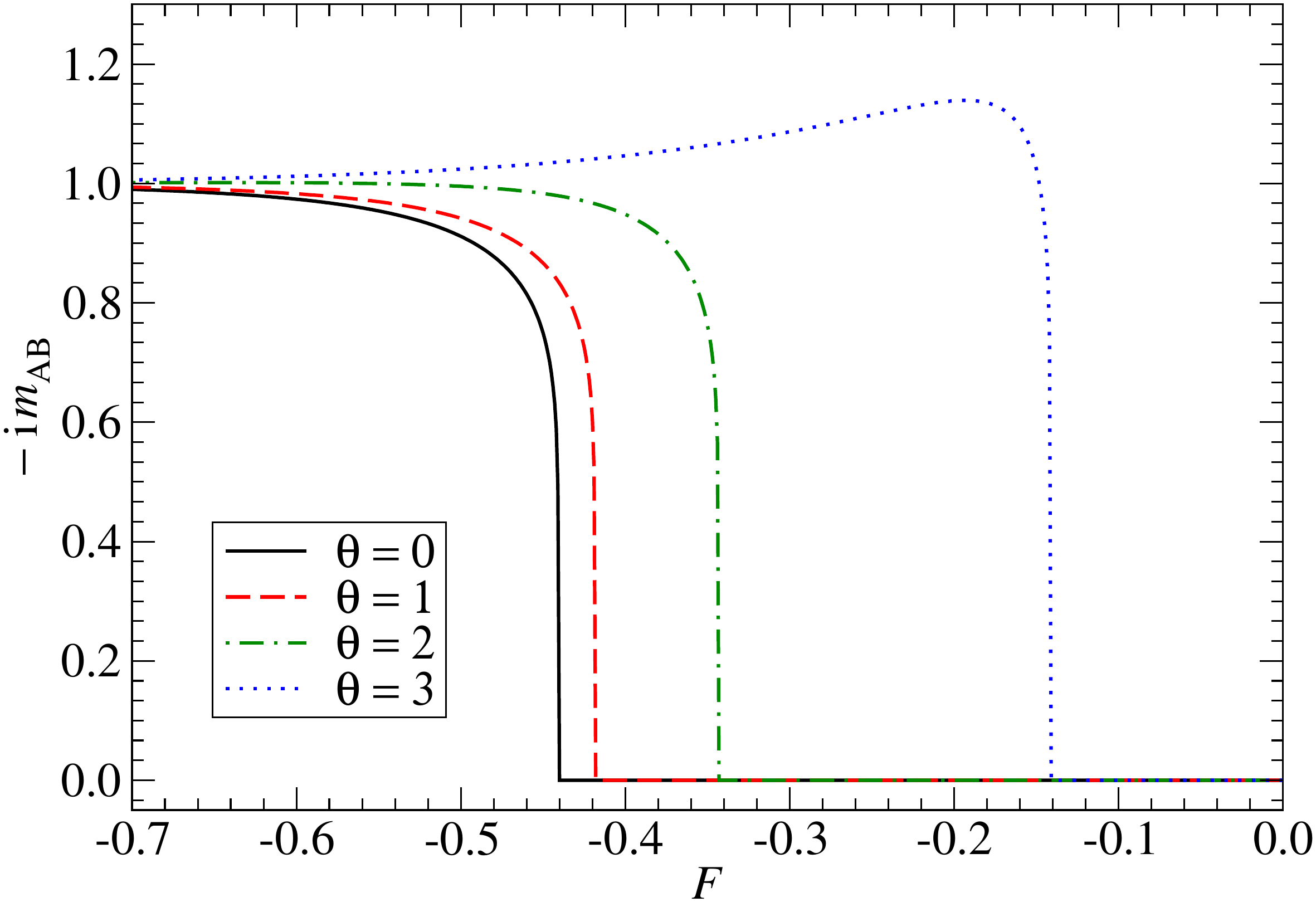}
\caption{Antiferromagnet:
The plot of the difference between sublattice magnetizations
(\ref{mminus}) $-{\rm i} m_{AB}$ versus the coupling $F$, for imaginary
fields $\theta=1,2,3$.
The spontaneous magnetization for the zero magnetic field $\theta=0$
is drawn for comparison.}
\label{fig:mminus} 
\end{center}
\end{figure}

Figure \ref{fig:mminus} shows the dependence of the magnetization
difference between the $A$ and $B$ sublattices, namely the real quantity
$-{\rm i} m_{AB}$, on the coupling $F$ for three values of
the imaginary field $\theta=1,2,3$; the spontaneous magnetization
for the zero magnetic field $\theta=0$ is presented as well.
The magnetization difference is zero above the critical coupling
$F_c$ and goes to $1$ for asymptotically large $F\to -\infty$.
The plot of the function $-{\rm i} m_{AB}(F)$ is non-monotonous
for $\theta=3$, it acquires a maximum larger than $1$.
This confirms that also the spontaneous magnetization difference between
two alternating sublattices can be larger than 1 for imaginary
magnetic fields.

\begin{figure}[t]
\begin{center}
\includegraphics[width=0.48\textwidth,clip]{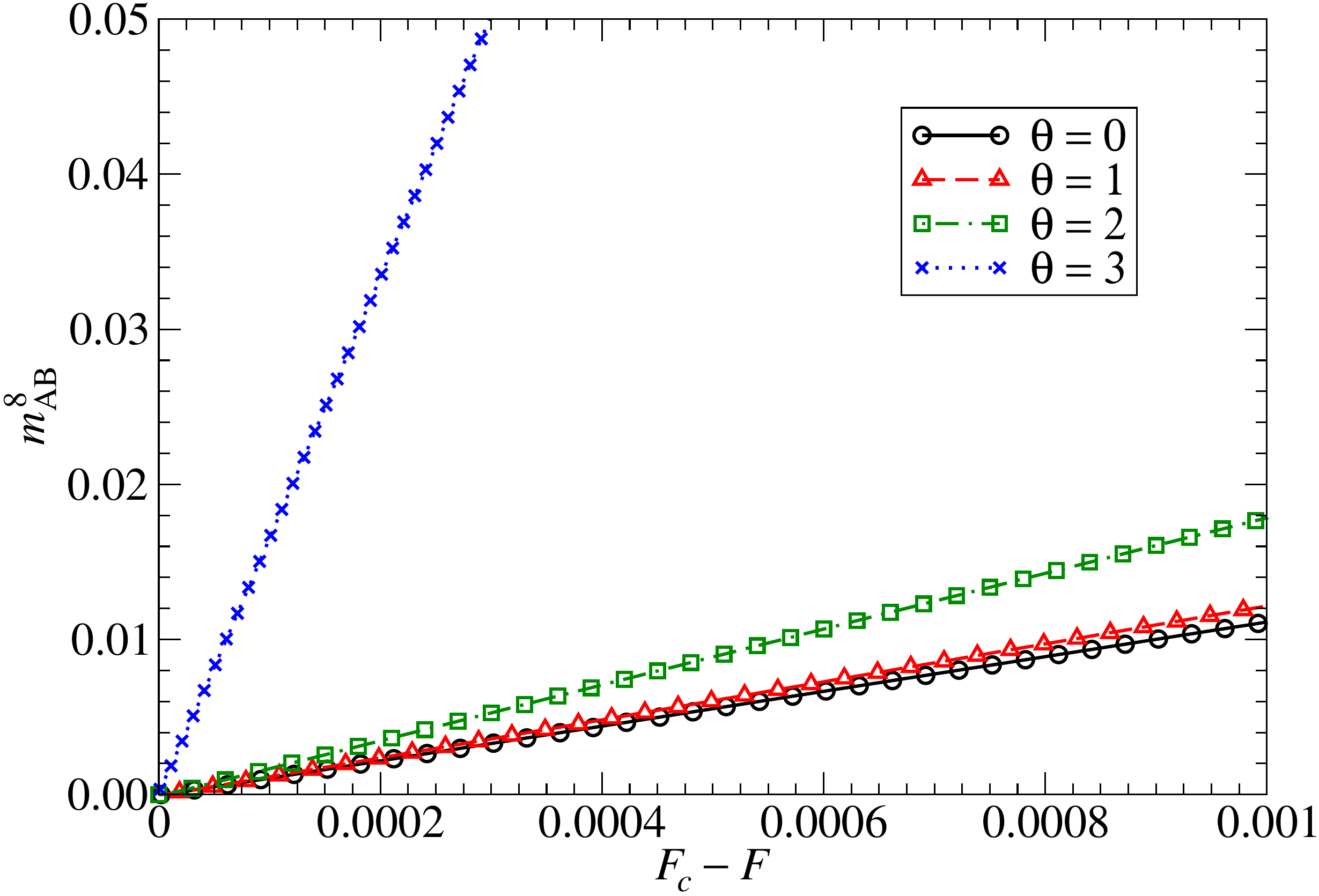}
\caption{Antiferromagnet:
The linear dependence of the 8th power of the difference
between sublattice magnetizations in the ordered phase on small deviations
from the critical coupling $F_c(\theta) - F$ for the zero magnetic
field $\theta=0$ and the values $\theta=1,2,3$ of the imaginary field.
The linear form of the plots indicates the uniformity of the critical
index $\beta=\frac{1}{8}$ along the line of critical points
when changing the parameter $\theta\in [0,\pi]$.}
\label{fig:beta} 
\end{center}
\end{figure}

The dependence of $m_{AB}^8$ on small deviations from the critical
coupling $F_c(\theta) - F$ for zero magnetic field $\theta=0$
and the imaginary fields $\theta=1,2,3$ is pictured in
Fig.~\ref{fig:beta}.
The linear form of the plots indicates that the critical exponent $\beta$
is constant along the whole curve of critical points $F_c(\theta)$,
equal to its zero-field Ising value $\frac{1}{8}$.
This behavior, which agrees with the universality hypothesis
\cite{Baxterbook,Samajbook}, was observed also for the 2D Ising
antiferromagnet in real non-zero magnetic fields \cite{Kolesik93}.

\begin{figure}[t]
\begin{center}
\includegraphics[width=0.48\textwidth,clip]{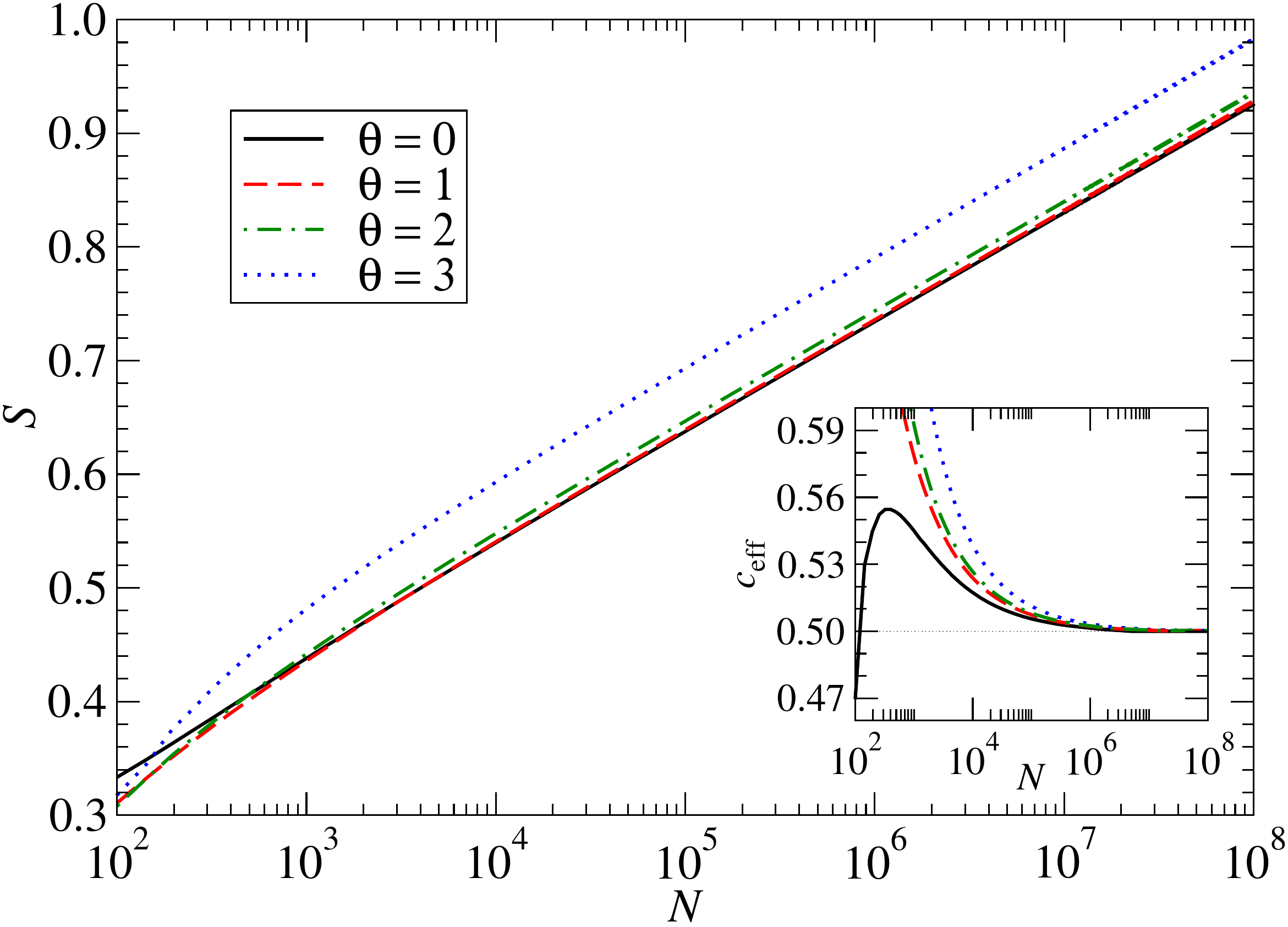}
\caption{Antiferromagnet:
The plot of the von-Neumann entropy $S$ versus the number of
sites of the square lattice $N$, in the logarithmic scale. 
The inset shows that with increasing $N$ the effective anomaly number
$c_{\rm eff}(N)$ (\ref{ceff}) tends to the Ising value $\frac{1}{2}$,
for the zero magnetic field $\theta=0$ as well as the values
$\theta=1,2,3$ of the imaginary magnetic field.}
\label{fig:c} 
\end{center}
\end{figure}

The plot of the von-Neumann entropy $S$ versus the number of sites $N$
of the square lattice is pictured in the logarithmic scale
in Fig.~\ref{fig:c}. 
It is evident that for large $N$ the entropy grows in accordance
with the expected asymptotic formula (\ref{entropyscale}). 
The inset documents the tendency of the effective anomaly number
$c_{\rm eff}(N)$ to the Ising value $\frac{1}{2}$ with increasing $N$,
for the zero field $\theta=0$ as well as for any value $\theta=1,2,3$
of the imaginary magnetic field.
This means that the presence of the imaginary magnetic field
does not change the universality class of the Ising antiferromagnet
and all critical exponents remain the same as those in the zero field.

\section{Numerical results for the 2D ferromagnet}
\label{section:resultsferro}

\begin{figure}[t]
\begin{center}
\includegraphics[width=0.48\textwidth,clip]{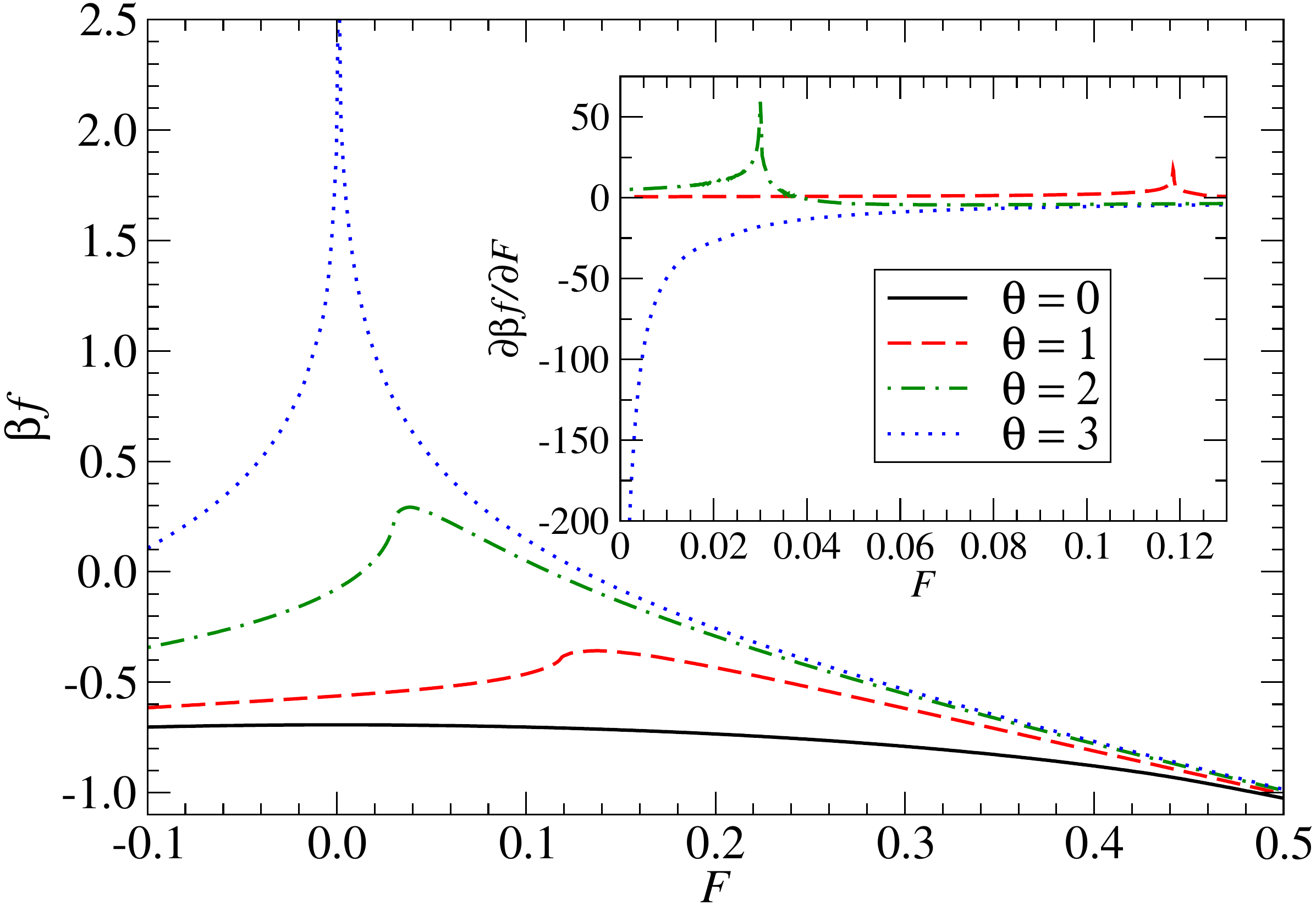}
\caption{Ferromagnet:
The free energy $\beta f$ as the function of the ferromagnetic
coupling $F>0$ for zero magnetic field $\theta=0$ and the values
$\theta=1,2,3$ of the imaginary field.
The inset shows the $F$-dependence of the first derivative of the free energy
with respect to $F$; the cusp divergence of the derivative
signalizes a first-order transition at the coupling $F^*(\theta)$.
The free energy is well defined in the high-temperature region $F<F^*(\theta)$,
the continuous plot of the free energy in the low-temperature region
$F>F^*(\theta)$ ignores the divergence of $\beta f$ at a dense set of
Yang-Lee zeros.}
\label{fig:firstorder} 
\end{center}
\end{figure}

Using the vertex representation of the 2D Ising ferromagnet on the
square lattice $(q=4)$ derived in Sec.~\ref{section:mappingb},
the dependence of the (dimensionless) free energy $\beta f$ on the
ferromagnetic coupling $F>0$ is pictured in Fig.~\ref{fig:firstorder}, 
for zero magnetic field $\theta=0$ and the values $\theta=1,2,3$
of the imaginary field.
The antiferromagnetic region of the couplings $F\in [-0.1,0]$ is
included to describe in detail the neighborhood of the point $F=0$.
While for the zero field the free energy as the function of $F$
decays monotonously, for the imaginary magnetic fields $\beta f$ first
grows in the region of small $F$ up to a maximum point and then decays
monotonously up to $F\to\infty$.
The curves are close to each other in the low-temperature
region, namely for large enough couplings $F\gtrapprox 0.5$.
The inset of Fig.~\ref{fig:firstorder} shows the $F$-dependence of the first
derivative of the free energy with respect to $F$; the cusp divergence of
the derivative signalizes a first-order phase transition
at the transient coupling $F^*(\theta)$.

To explain our numerical data for the free energy in more detail, we recall
that, in analogy with the 1D version of the model, the free energy is
expected to be well defined in the high-temperature region $F<F^*(\theta)$. 
In this region, being sufficiently far away from $F^*(\theta)$ we calculate
the free energy at equidistant points on the $F$-axis with the step
$\Delta F = 0.01$.
When the free energy starts to vary substantially, i.e. when one is close
to the transient point $F^*(\theta)$, in order to describe correctly
the neighborhood of $F^*(\theta)$ the equidistant step is changed to
the smaller one $\Delta F = 0.0001$. 
Passing through the transient point $F^*(\theta)$, the free energy changes
smoothly once again and one returns to the previous step $\Delta F = 0.01$.
It stands to reason that in the low-temperature region $F>F^*(\theta)$
there exist the problematic dense set of Yang-Lee zeros of the partition
function at which the free energy per site blows up to infinity.
As is evident from Fig.~\ref{fig:firstorder}, our choice of the rational
equidistant points on the $F$-axis does not involve Yang-Lee zeros and
the numerical plot of the free energy versus $F$ looks to be continuous
This mathematical peculiarity of limited physical interest 
is in close analogy with the 1D version of the Ising ferromagnet and
we suggest that one has to be precisely at a Yang-Lee zero to observe
the divergence of the free energy.

\begin{figure}[t]
\begin{center}
\includegraphics[width=0.48\textwidth,clip]{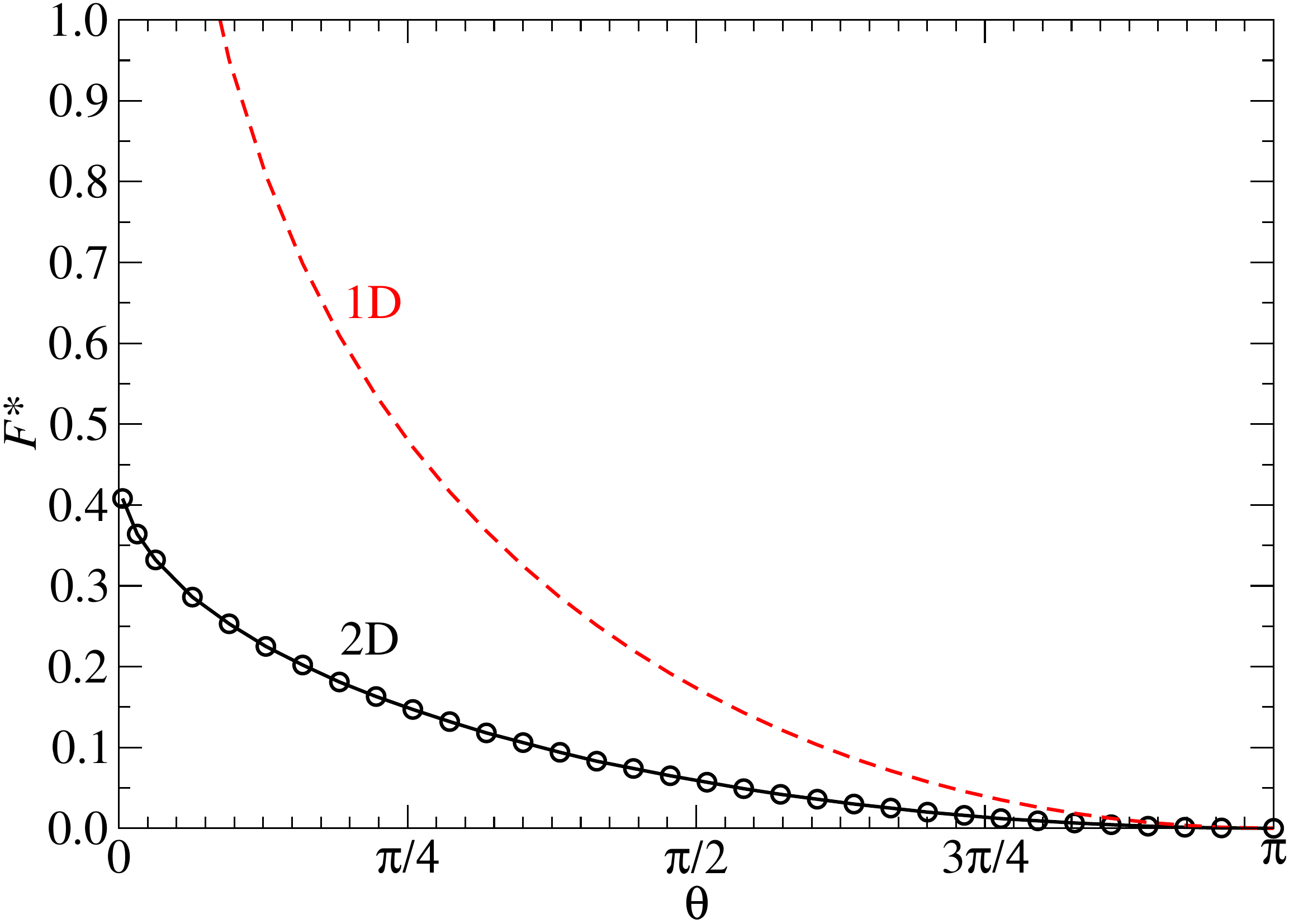}
\caption{Ferromagnet:
The transition coupling $F^*$ as the function of the
imaginary magnetic field $\theta$.
The analytic 1D result (\ref{defFstar}) is drawn by the dashed curve,
the numerical 2D data are represented by open circles.}
\label{fig:Fstar} 
\end{center}
\end{figure}

The dependences of the first-order transition coupling $F^*$ on the imaginary
magnetic field $\theta$ in 1D and 2D are pictured in Fig.~\ref{fig:Fstar}.
The analytic 1D result (\ref{defFstar}) is drawn by the dashed curve.
It is seen that as $\theta\to 0$ the coupling $F^*\to\infty$ which
is in agreement with the fact that for the Ising ferromagnet in zero
magnetic field  there is neither first-order phase transition nor
the divergence of the magnetization.
The numerical 2D data for the dependence $F^*(\theta)$ are represented
in Fig.~\ref{fig:Fstar} by open circles.
For each $\theta$, the value of the 2D $F^*(\theta)$ is always smaller than
the one in 1D.
The limiting $\theta\to 0^+$ value of $F(\theta)$ is a finite number.
The 2D Ising ferromagnet at the strictly zero magnetic field $\theta=0$
exhibits no first-oder phase transition and, consequently, $F^*(0)$
does not exists.

\begin{figure}[t]
\begin{center}
\includegraphics[width=0.48\textwidth,clip]{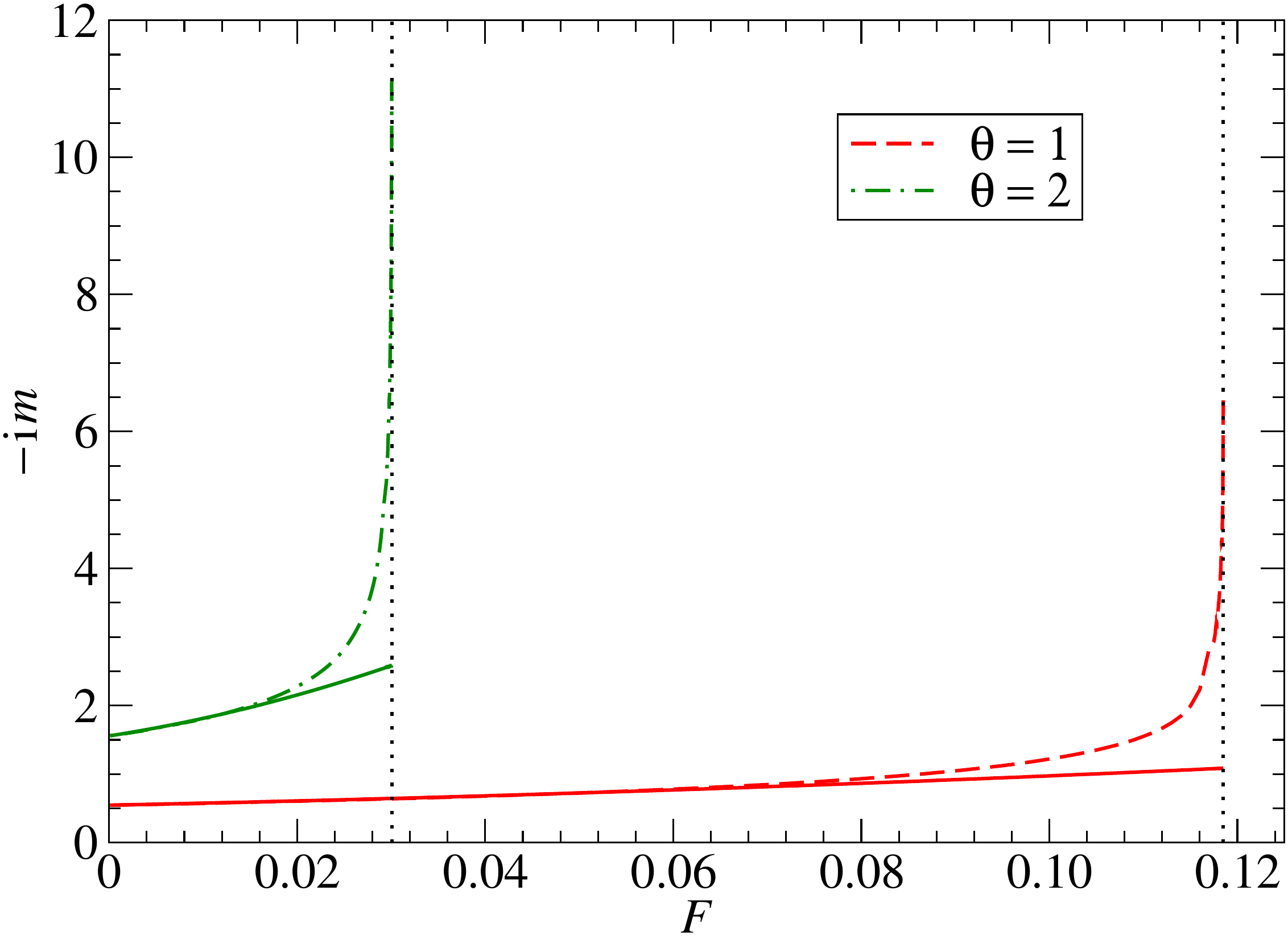}
\caption{Ferromagnet:
The divergence of the magnetization $-{\rm i} m$ for the 2D Ising ferromagnet
in an imaginary field as the coupling constant $F$ approaches
the transition coupling $F^*(\theta)$ (vertical dotted lines) from below;
the imaginary magnetic field $\theta=1$ (dashed curve) and $\theta=2$
(dash-dotted curve).
The dependences of $-{\rm i} m$ versus $F$ yielded by the asymptotic
$F\to 0$ formula (\ref{asymptotic}) are depicted by the solid lines.}
\label{fig:magndiverg} 
\end{center}
\end{figure}

The divergence of the magnetization $-{\rm i} m$ when the coupling constant
$F$ approaches the transition coupling $F^*(\theta)$ (vertical dotted lines)
from below (i.e., from the high-temperature region) is represented
in Fig.~\ref{fig:magndiverg}.
The dashed curve corresponds to the imaginary magnetic field $\theta=1$
and the dash-dotted curve to $\theta=2$.
Data for the imaginary magnetic field $\theta=3$ are omitted since
$F^*(3)$ is very close to zero which causes numerical instabilities in
the calculation of the magnetization plot.
The values of $F^*(\theta)$ obtained in this way coincide with a high
accuracy with the previous ones obtained from the divergence of the
first derivative of the free energy, see Fig.~\ref{fig:Fstar}.
The high-temperature expansion of the magnetization for the Ising model on
the square lattice in a magnetic field in powers of the nearest-neighbor
coupling is written in Eq. (1.8.7) of monograph \cite{Baxterbook}.
Inserting there imaginary field, one obtains
\begin{eqnarray}
- {\rm i} m & = & \tan\left( \frac{\theta}{2} \right) +
\frac{4\tan\left( \frac{\theta}{2} \right)}{
\cos^2\left(\frac{\theta}{2}\right)} F \nonumber \\
& & + \frac{4\tan\left( \frac{\theta}{2} \right)}{
\cos^2\left(\frac{\theta}{2}\right)}
\left[ 3 + 7 \frac{\sin^2\left(\frac{\theta}{2}\right)}{\cos^2
\left(\frac{\theta}{2}\right)} \right] F^2 + {\cal O}(F^3) . \phantom{aaa}  
\label{asymptotic}
\end{eqnarray}  
The dependences of $-{\rm i} m$ on $F$ yielded by this asymptotic
relation are depicted for the imaginary fields $\theta=1, 2$ in
Fig.~\ref{fig:magndiverg} by the solid lines for comparison.
When $F>F^*(\theta)$, the thermodynamic limit of the magnetization
changes chaotically with the system size (not shown in the figure)
and so it is ill-defined like in 1D.

\section{Conclusion} \label{section:conclusion}
The Ising model in a pure imaginary magnetic field exhibits a severe sign
problem.
The Boltzmann weight of a configuration of spins on the lattice is
a complex number which prevents from application of standard methods
in equilibrium statistical mechanics.
We avoid this problem by mapping the considered Ising model on the square
lattice onto the symmetric vertex model (with the permutation symmetry of
local vertex weights) formulated on the same lattice structure
in Sec.~\ref{section:mapping}.
The mapping depends on whether the Ising nearest-neighbor couplings
are antiferromagnetic (Sec.~\ref{section:mappinga}) or ferromagnetic
(Sec.~\ref{section:mappingb}).
The local vertex weights of the symmetric vertex model are
real (positive or negative) numbers.
This fact permits us to apply the accurate numerical CTMRG and HOTRG
methods based on the renormalization of the density matrix.
Another potential application of the mapping onto the symmetric
vertex model with real vertex weights is to search for the Yang-Lee zeros of
the Ising partition function on finite lattices, e.g., with periodic boundary
conditions.

The numerical results for the 2D antiferromagnet in an imaginary magnetic
field are presented in Sec.~\ref{section:resultsantiferro}.
The curve of critical points separating the ordered antiferromagnetic
and the disordered paramagnetic phases of the model in Fig.~\ref{fig:critF}
is estimated with the numerical precision of order $0.2\%$ which
substantially overcomes the accuracy of other methods
\cite{Matveev08,Azcoiti17}.
Data for the magnetization difference between two interwoven sublattices
as the function of the coupling $F$ are pictured in Fig.~\ref{fig:mminus}.
It is seen that for the sufficiently high imaginary field $\theta=3$ 
the spontaneous difference between the sublattice magnetizations turns out
to be larger than one in an interval of the couplings $F$.
As concerns the critical properties, there is numerical evidence that
the critical exponent $\beta$ (Fig.~\ref{fig:beta}) and the
anomaly number $c$ (Fig.~\ref{fig:c}) do not depend on the strength of
the imaginary magnetic field and are equal to the zero-field Ising
values $\beta=\frac{1}{8}$ and $c=\frac{1}{2}$.

The phase properties of the 2D ferromagnetic Ising model in an imaginary
magnetic field, studied in Sec.~\ref{section:resultsferro}, are qualitatively
similar to those of its 1D version (Sec.~\ref{1Dferro}).
In particular, there is a first-order transition coupling $F^*(\theta)$
at which both the first derivative of the free energy with respect to
the coupling and the magnetization diverge when approaching to
$F^*(\theta)$ from the high-temperature side, $F\to {F^*}^-$.
The free energy and the magnetization per site are well defined
in the high-temperature region $F<F^*(\theta)$.  
In the low-temperature region $F>F^*(\theta)$, the free energy blows up at
couplings which correspond to the Yang-Lee zeros of the partition function.
As is evident from Fig.~\ref{fig:firstorder}, our choice of equidistant
rational points on the $F$-axis does not involve Yang-Lee zeros and
the numerical plot of the free energy versus $F$ looks to be smooth.
This mathematical curiosity of limited physical interest was explained on
the exactly solvable 1D ferromagnet in Sec. \ref{1Dferro} based on plausible
arguments.
The magnetization depends chaotically on the system size for $F>F^*(\theta)$
and therefore it is ill-defined for both 1D and 2D.
The only fundamental difference between 1D and 2D comes from
Fig.~\ref{fig:Fstar}: while the zero-field $\theta\to 0$
limit of $F^*(\theta)$ goes continuously to the expected value $\infty$
in 1D, it approaches to a finite value in 2D and does not exist at
the strictly zero field $\theta=0$.

\begin{acknowledgments}
The support received from the Grants VEGA Nos. 2/0123/19 and 2/0092/21,
Joint Research Project SAS-MOST 108-2112-M-002-020-MY3 and Project
APVV-20-0150 is acknowledged. 
\end{acknowledgments}

\section*{Appendix} \label{section:appendix}
Let the standard impurity tensor $T_m$ be given by the product of the
vertex weights, see (\ref{permutation}).
The standard impurity tensor is used to evaluate the magnetization
according to $m={\rm Tr}\left( T_m \right)$.
For a spatially homogeneous system on the square lattice with
the coordination number $q=4$, the standard impurity tensor reads
\begin{eqnarray}
T_m(\sigma^{~}_1,\sigma^{~}_2,\sigma^{~}_3,\sigma^{~}_4)
& = & \sum_{\sigma'_1,\sigma'_2,\sigma'_3,\sigma'_4=\pm} 
v_m(\sigma'_1,\sigma'_2,\sigma'_3,\sigma'_4) \nonumber \\ 
& & \times W_{\sigma'_1,\sigma_1} W_{\sigma'_2,\sigma_2} 
W_{\sigma'_3,\sigma_3} W_{\sigma'_4,\sigma_4} \phantom{aaa}
\end{eqnarray} 
with the vertex tensor $v_m$ being
\begin{equation}
v_m(\sigma_1,\sigma_2,\sigma_3,\sigma_4) = {\rm e}^{{\rm i}\frac{\theta}{2}}
\prod_{j=1}^4 \delta(\sigma_j,+) - {\rm e}^{-{\rm i}\frac{\theta}{2}}
\prod_{j=1}^4 \delta(\sigma_j,-) .
\end{equation}

For the inhomogeneous system, however, the sublattices magnetizations
$m_A$ and $m_B$ are not identical in the symmetry broken state anymore.
Then, the impurity tensor has to be redefined to describe the nonzero
difference of the magnetization
$m_{AB} = m_A - m_B = \max \left( T_{AB} \right)$.
We, therefore, consider the maximal absolute value of the extended
impurity tensor $T_{AB}$ (rather than its trace).
The construction of $T_{AB}$ was carried out by means of additional
four extended impurity tensors  $T_1^A$, $T_2^A$, $T_1^B$, and $T_2^B$,
such that
\begin{equation}
T_{AB} = T_1^A + T_2^A - T_1^B - T_2^B.
\end{equation}
These extended impurity tensors have doubled ranks, because the degrees
of freedom on the edges of the tensors are squared, while the coordination
number still remains unchanged ($q=4$).
The four tensors satisfy the relations
\begin{eqnarray}
T_1^A (\{\sigma_1\bar\sigma_1\},\{\sigma_2\bar\sigma_2\},
\{\sigma_3\bar\sigma_3\},\{\sigma_4\bar\sigma_4\}) =\nonumber \\
\prod_{\sigma'_1,\sigma'_2,\sigma'_3,\sigma'_4}
T_{m}(\bar\sigma_1,\sigma_2,\sigma'_1,\sigma'_4)
w(\sigma'_1,\bar\sigma_2,\bar\sigma_3,\sigma'_2)
\nonumber \\ \times
w(\sigma'_3,\sigma'_2,\sigma_3,\bar\sigma_4)
w(\sigma_1,\sigma'_4,\sigma'_3,\sigma_4) ,
\end{eqnarray}
\begin{eqnarray}
T_2^A (\{\sigma_1\bar\sigma_1\},\{\sigma_2\bar\sigma_2\},
\{\sigma_3\bar\sigma_3\},\{\sigma_4\bar\sigma_4\}) =\nonumber \\
\prod_{\sigma'_1,\sigma'_2,\sigma'_3,\sigma'_4}
w(\bar\sigma_1,\sigma_2,\sigma'_1,\sigma'_4)
w(\sigma'_1,\bar\sigma_2,\bar\sigma_3,\sigma'_2)
\nonumber \\ \times
w(\sigma'_3,\sigma'_2,\sigma_3,\bar\sigma_4)
T_m(\sigma_1,\sigma'_4,\sigma'_3,\sigma_4) ,
\end{eqnarray}
\begin{eqnarray}
T_1^B (\{\sigma_1\bar\sigma_1\},\{\sigma_2\bar\sigma_2\},
\{\sigma_3\bar\sigma_3\},\{\sigma_4\bar\sigma_4\}) =\nonumber \\
\prod_{\sigma'_1,\sigma'_2,\sigma'_3,\sigma'_4}
w(\bar\sigma_1,\sigma_2,\sigma'_1,\sigma'_4)
T_m(\sigma'_1,\bar\sigma_2,\bar\sigma_3,\sigma'_2) \nonumber \\ \times
w(\sigma'_3,\sigma'_2,\sigma_3,\bar\sigma_4)
w(\sigma_1,\sigma'_4,\sigma'_3,\sigma_4) ,
\end{eqnarray}
\begin{eqnarray}
T_2^B (\{\sigma_1\bar\sigma_1\},\{\sigma_2\bar\sigma_2\},
\{\sigma_3\bar\sigma_3\},\{\sigma_4\bar\sigma_4\}) =\nonumber \\
\prod_{\sigma'_1,\sigma'_2,\sigma'_3,\sigma'_4}
w(\bar\sigma_1,\sigma_2,\sigma'_1,\sigma'_4)
w(\sigma'_1,\bar\sigma_2,\bar\sigma_3,\sigma'_2)
\nonumber \\ \times
T_m(\sigma'_3,\sigma'_2,\sigma_3,\bar\sigma_4)
w(\sigma_1,\sigma'_4,\sigma'_3,\sigma_4) .
\end{eqnarray}

Within the HOTRG method, the impurity tensors iteratively expand into
the doubled ranks and renormalize back to their original ranks.
They represent linear combinations of all possible positions of
the magnetic tensor (with the appropriate sign) divided by number
of the combinations taken.


\begin{thebibliography}{10}

\bibitem{Uzelac80} K. Uzelac, R. Jullien, and P. Pfeuty,
One-dimensional transverse-field Ising model in a complex longitudinal
field from a real-space renormalization-group method at $T=0$,
Phys. Rev. B {\bf 22}, 436 (1980).
  
\bibitem{Gehlen91} G. von Gehlen,
Critical and off-critical analysis of the Ising quantum chain in an
imaginary field,  
J. Phys. A: Math. Gen. {\bf 24}, 5371 (1991).
  
\bibitem{Deguchi09} T. Deguchi and P. K. Ghosh,
The exactly solvable quasi-Hermitian transverse Ising model,
J. Phys. A: Math. Theor. {\bf 42}, 475208 (2009).
  
\bibitem{Matveev95} V. Matveev and R. Shrock,
Complex-temperature properties of the 2D Ising model with
$\beta H = \pm {\rm i}\pi/2$,
J. Phys. A: Math. Gen. {\bf 28}, 4859 (1995).
  
\bibitem{Matveev96} V. Matveev and R. Shrock,
Complex-temperature properties of the two-dimensional Ising model for
nonzero magnetic field,
Phys. Rev. E {\bf 53}, 254 (1996).

\bibitem{Matveev08} V. Matveev and R. Shrock,
On properties of the Ising model for complex energy/temperature and
magnetic field,
J. Phys. A: Math. Theor. {\bf 41}, 135002 (2008).

\bibitem{Kim04} S.-Y. Kim,
Yang-Lee zeros of the antiferromagnetic Ising model,
Phys. Rev. Lett. {\bf 93}, 130604 (2004).  

\bibitem{Muller77} E. M\"uller-Hartmann and J. Zittartz,
Interface free energy and transition temperature of the square-lattice
Ising antiferromagnet at finite magnetic field,
Z. Phys. B {\bf 27}, 261 (1977).

\bibitem{Wu89} F. Y. Wu, X. N. Wu, and H. W. J. Bl\"{o}te,
Critical frontier of the antiferromagnetic Ising model in a magnetic
field: The honeycomb lattice,  
Phys. Rev. Lett. {\bf 62}, 2773 (1989).  

\bibitem{Kolesik93} M. Koles\'{\i}k and L. \v{S}amaj,
New variational series expansions for lattice models,
J. Phys. I France {\bf 3}, 93 (1993). 

\bibitem{Azcoiti17} V. Azcoiti, G. Di Carlo, E. Follana, and E. Royo-Amondarain,
Antiferromagnetic Ising model in an imaginary field,
Phys. Rev. E {\bf 96}, 032114 (2017).

\bibitem{Azcoiti11} V. Azcoiti, E. Follana, and A. Vaquero,
Progress in numerical simulations of systems with a $\theta$-vacuum like term:
The two and three-dimensional Ising model within an imaginary magnetic field,   
Nucl. Phys. B {\bf 851} [FS], 420 (2011).

\bibitem{Azcoiti00} V. Azcoiti, V. Laliena, and A. Galante,
QCD with a $\theta$-vacuum term: A complex system with a simple complex
action, in: Proceedings of the International Workshop on Non-Perturbative
methods and Lattice QCD, Guangzhou, China, p.161 (2000). 

\bibitem{Onsager44} L. Onsager,
Crystal statistics. I. A two-dimensional model with an order disorder
transition,  
Phys. Rev. {\bf 65}, 117 (1944).

\bibitem{Lee52} T. D. Lee and C. N. Yang,
Statistical theory of equations of state and phase transitions.
II. Lattice gas and Ising model,  
Phys. Rev. {\bf 87}, 410 (1952).   

\bibitem{Azcoiti99} V. Azcoiti and A. Galante,
Parity and CT realization in QCD,
Phys. Rev. Lett. {\bf 83}, 1518 (1999).     

\bibitem{Gradshteyn} I. S. Gradshteyn and I. M. Ryzhik,
{\it Table of Integrals, Series, and Products,} 6th ed.
(Academic Press, London, 2000).  

\bibitem{EM} M. Abramowitz and I. A. Stegun,
{\it Handbook of Mathematical Functions with Formulas, Graphs, and
Mathematical Tables,} 9th ed.
(Dover, New York, 1972).

\bibitem{Samaj92} L. \v{S}amaj and M. Koles\'{\i}k,
Mapping of the symmetric vertex model onto the Ising model for
an arbitrary lattice coordination,
Physica A {\bf 182}, 455 (1992).

\bibitem{Wegner73} F. J. Wegner,
A transformation including the weak-graph theorem and the duality
transformation,
Physica {\bf 68}, 570 (1973).
  
\bibitem{Nagle68} J. F. Nagle,
Weak-graph method for obtaining formal series expansions for lattice
statistical problems,
J. Math. Phys. {\bf 9}, 1007 (1968).  

\bibitem{White92} S. R. White,
Density matrix formulation for quantum renormalization groups,
Phys. Rev. Lett. {\bf 69}, 2863 (1992).

\bibitem{White93} S. R. White,
Density-matrix algorithms for quantum renormalization groups,
Phys. Rev. B {\bf 48}, 10345 (1993).  

\bibitem{Schollwock05} U. Schollw\"{o}ck,
The density-matrix renormalization group,
Rev. Mod. Phys. {\bf 77}, 259 (2005).  

\bibitem{Nishino96} T. Nishino and K. Okunishi,
Corner transfer matrix renormalization group method,
J. Phys. Soc. Jpn. {\bf 65}, 891 (1996).

\bibitem{Nishino97} T. Nishino and K. Okunishi,
Corner transfer matrix algorithm for classical renormalization group,
J. Phys. Soc. Jpn. {\bf 66}, 3040 (1997).

\bibitem{Ueda05} K. Ueda, R. Otani, Y. Nishio, A. Gendiar, and T. Nishino,
Critical point of a symmetric vertex model,
J. Phys. Soc. Jpn. {\bf 74}, 1871 (2005).  

\bibitem{Xie12} Z. Y. Xie, J. Chen, M. P. Qin, J. W. Zhu, L. P. Yang, and
T. Xiang,
Coarse-graining renormalization by higher-order singular value decomposition,
Phys. Rev. B {\bf 86}, 045139 (2012).

\bibitem{Baxterbook} R. J. Baxter,
{\it Exactly Solved Models in Statistical Mechanics}, 3rd ed.
(Dover Publications, London, 2007).

\bibitem{Calabrese04} P. Calabrese and J. Cardy,
Entanglement entropy and quantum field theory,
J. Stat. Mech. P06002 (2004).  

\bibitem{Ercolessi10} E. Ercolessi, S. Evangelisti, and F. Ravanini,
Exact entanglement entropy of the XYZ model and its sine-Gordon limit,
Phys. Lett. A {\bf 374}, 2101 (2010).  

\bibitem{Samajbook} L. \v{S}amaj and Z. Bajnok,
{\it Introduction to the Statistical Physics of Integrable Many-body Systems},
(Cambridge Univ. Press, Cambridge, 2013).

\end{thebibliography}
\end{document}